\documentclass[aps,prb,twocolumn,superscriptaddress,showpacs]{revtex4}

\usepackage{graphicx}
\newcommand{\be}{\begin{equation}}
\newcommand{\ee}{\end{equation}}
\newcommand{\bea}{\begin{eqnarray}}
\newcommand{\eea}{\end{eqnarray}}

\begin{document}

\title{Magnetic field induced phase transitions in spin ladders with ferromagnetic legs}

\author{ T. Vekua}
\affiliation{Institut f\"{u}r Theoretische Physik,
Universit\"{a}t Hannover, 30167 Hannover, Germany}

\author{G.I. Japaridze}
\affiliation{Andronikashvili Institute of Physics, Tamarashvili 6, 0177, 
Tbilisi, Georgia}

\author{H.-J. Mikeska}
\affiliation{Institut f\"{u}r Theoretische Physik,
Universit\"{a}t Hannover, 30167 Hannover, Germany}

\begin{abstract} 
  We study the phase diagram of spin ladders with ferromagnetic legs under the
  influence of a symmetry breaking magnetic field in the weak coupling
  effective field theory by bosonization. For antiferromagnetic interleg
  coupling we identify two phase transitions introduced by the external
  magnetic field. In order to establish the universality of the phases we
  supplement the bosonization approach by results from a strong coupling
  (rung) expansion and from spin wave analysis.
\end{abstract}

\maketitle

\section{Introduction}

Recently there has been considerable interest in the study of magnetic
field-induced effects in low-dimensional quantum spin systems, in particular
devoted to the critical properties of spin $S=1/2$ {\em isotropic
  antiferromagnetic} two-leg ladders in an external magnetic field.
In parallel, there was remarkable progress in recent years in the
fabrication of such ladder compounds \cite{RiceDagotto}. Since
antiferromagnetic two-leg ladder systems with $S = 1/2$ have a gap in
the spin excitation spectrum, they reveal an extremely rich behavior,
dominated by quantum effects, in the presence of a magnetic
field. These quantum phase transitions were intensively investigated
both theoretically, using different analytical and numerical
techniques
\cite{ChitraGiamarchi,Honecker,Totsuka,Usami,GiamarchiTsvelik,Mikeska1,Wang1,Haas,Wang2},
and also experimentally
\cite{Chaboussant1,Chaboussant2,Chaboussant3,Exp4,Exp5,Exp6}.

Ladder models with {\em ferromagnetic legs} have been much less
studied, although they exhibit many interesting aspects
\cite{Schulz1,KolezhukMikeska,Vekua}. It is true that up to now no
materials are available which realize these models. However, from the
theoretical point of view these systems are extremely interesting,
since they open up a new large class of systems for the study of
complicated quantum behavior, unexpected in more conventional spin
systems. The variety of possibilities is seen already from
Fig.~\ref{fig:ladder}: here the ground state phase diagram
\cite{Vekua} of a two-leg ferromagnetic ladder is presented in the
variables intraleg exchange anisotropy ($\Delta$) and (isotropic)
interleg coupling ($J_{\perp}$). The ground state phase diagram
contains, besides the fully gapped rung-singlet and Haldane phases
(commonly known from the case of antiferromagnetic ladders
\cite{Schulz1,Nersesyan96}), the spin-liquid phase with easy-plane
anisotropy ($XY1$), the ferromagnetic and the stripe-ferromagnetic
phases which are realized only in the case of ferromagnetic legs ($0
\leq \Delta \leq 1$).

In this paper we study the effect of an external magnetic field on the
phase diagram of this system. In particular, we focus our attention on
the study of new field induced effects in the case of the easy-plane $XY1$
phase and in the rung-singlet phase in the vicinity of the single
chain ferromagnetic instability point $\Delta=1$.

The effect of the uniform magnetic field, applied parallel to the anisotropy 
($z$) axis, on the ground state properties of the two-leg
ladder systems is known from the investigation by Schulz \cite{Schulz1}
for the case of the corresponding spin $S=1$ Heisenberg chain model:
In the case of the gapless $XY1$ phase an external magnetic field
leads to the appearance of a magnetization in $z$ direction for
arbitrary small magnetic fields. In the case of the gapped
rung-singlet phase the magnetization appears only at a finite critical
value of the magnetic field which is equal to the spin gap
\cite{Schulz1,ChitraGiamarchi}. This behavior is generic for the spin
gapped $U(1)$ symmetric systems in a magnetic field: The magnetic
field leaves the in-plane rotational invariance unchanged \cite{JNW}
and the transition belongs to the universality class of the
commensurate-incommensurate (C-IC) transitions
\cite{Japaridze,Pokrovsky}.

On the other side, for the $U(1)$ symmetric phase such as the $XY1$
phase, the effect of a uniform {\em transverse field} is highly
nontrivial. In the case of classical anisotropic spin
chains this effect has been studied more than two decades ago
\cite{Mikeska}. However, in the case of the antiferromagnetic $XXZ$
quantum chain this problem is still the subject of intensive recent
studies \cite{Hieida,Ovchinnikov,Essler03,DuttaSen}.

In this paper, we study the effect of a {\em uniform transverse}
magnetic field on the ground state phase diagram of a two-leg ladder
with {\em anisotropic, ferromagnetically interacting} legs coupled by
antiferromagnetic interleg exchange.

The outline of the paper is as follows: In section II we review the
model and its bosonized version in the continuum limit. We discuss the
phases and phase transitions emerging from the XY1 and rung-singlet
phases in a transverse magnetic field in sections III (in the weak
coupling approach) and IV (in the limit of strong rung exchange). We
shortly summarize our results in section V. In appendix A we present
the spin-wave approach to locate the ferromagnetic transition line
starting from the saturated phase.

\section{Model}

Here we present a brief introduction to the model and in particular to its
bosonized version in the continuum limit. The Hamiltonian we consider is given
by:
\begin{equation}
\label{Hamiltonian}
H = H_{leg}^{(1)} + H_{leg}^{(2)} +  H_{\perp}\, ,
\ee
where the Hamiltonian for leg $\alpha$ is
\bea\label{FerroLadderHamiltonian}
H_{leg}^{(\alpha)} =& - & J \sum_{j=1}^N 
\Big( S^{x}_{\alpha,j}S^{x}_{\alpha,j+1} +
S^{y}_{\alpha,j}S^{y}_{\alpha,j+1} \nonumber\\ 
&+& \Delta \,S^{z}_{\alpha,j}S^{z}_{\alpha,j+1}\Big) 
- {h^{ext}} \sum_{j=1}^N S^{x}_{\alpha,j}\, ,
\eea
and the interleg coupling is given by
\be\label{InterLegCoup}
H_{\bot} = J_{\bot} \sum_{j=1}^N \vec S_{j,1} \vec S_{j,2}\, .
\ee
Here $S^{x,y,z}_{\alpha,j}$ are spin $S=1/2$ operators on the $j$-th rung, 
and the index $\alpha=1,2$ denotes the ladder legs. The intraleg coupling 
constant is ferromagnetic, $J>0$, and therefore the limiting case of 
{\em isotropic ferromagnetic} legs corresponds to ${\Delta}=1$. We will 
restrict ourselves to the case $0\leq {\Delta} \leq 1$.

\begin{figure}[tb]
\includegraphics[width=80mm]{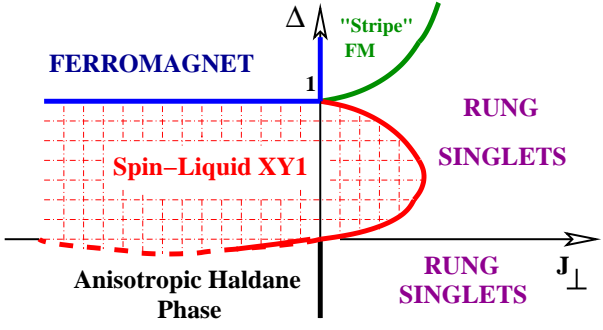}
\caption{\label{fig:ladder} Schematic picture of the ground state phase 
diagram of the two-leg ladder in the variables intraleg exchange 
anisotropy ($\Delta$) and isotropic interleg coupling ($J_{\perp}$).
The limiting case of {\em isotropic ferromagnetic} legs 
corresponds to ${\Delta}=1$.}
\end{figure}
We use the following bosonization expressions for spin operators, 
adapted to the case of ferromagnetic exchange (for details see\cite{Vekua}):
\begin{eqnarray}
S_{j,\alpha}^{x} &\simeq & \frac{{\it c}}{\sqrt{2\pi}} 
:\cos\sqrt{\frac{\pi}{K}} \theta_{\alpha}: \,\nonumber\\ 
+ &(-1)^j & \, \frac{{\it ib }}{\sqrt{2\pi}}
:\sin\sqrt{4\pi K}\phi_{\alpha} \sin \sqrt{\frac{\pi}{K}} \theta_{\alpha} :\, ,
\label{bosforSx}\\
S_{j,\alpha}^{y}  &\simeq &  \frac{{\it  c}}{\sqrt{2\pi}}
:\sin\sqrt{\frac{\pi}{K}} \theta_{\alpha}:\, \nonumber\\ 
-&(-1)^j&\, \frac{{\it ib}}{\sqrt{2\pi}} 
:\sin\sqrt{4\pi K}\phi_{\alpha}\cos\sqrt{\frac{\pi}{K}} \theta_{\alpha}:\, ,
\label{bosforSy}\\
S_{j,\alpha}^{z} &=&  \sqrt{\frac{K}{\pi}} \partial_x \phi_{\alpha}\, + \nonumber\\ 
&(-1)^j&  \,
 \frac{{\it a}}{\pi} :\sin\sqrt{4\pi K}\phi_{\alpha} (x): \, .
\label{bosforSz}
\end{eqnarray}
Here, $\phi(x)$ and $\theta(x)$ are dual bosonic fields, $\partial_t \phi =
u \partial_x \theta$ and $K$ is the Luttinger liquid parameter
\begin{equation}
K = \frac{\pi}{2\arccos\Delta}\, .
\end{equation}

Now we introduce the symmetric and antisymmetric combinations of the
bosonic fields $\phi_{\pm}=\sqrt{1/2}\left(\phi_1 \pm \phi_2\right)$, 
$\theta_{\pm}= \sqrt{1/2}\left(\theta_1 \pm \theta_2\right)$ and after 
rescaling these fields we obtain finally as effective 
bosonic Hamiltonian
\be\label{EffectiveHamiltonian}
{\cal H}= {\cal H}^{+} + {\cal H}^{-} + {\cal H}^{\pm}\, ,   
\ee
where
\begin{eqnarray}
{\cal H}^{+} & =& {u_{+} \over 2} [(\partial_x \theta_{+})^{2} + 
(\partial_x \phi_{+})^2] \nonumber\\
&+& M_{+} \cos\sqrt {8 \pi K_{+}}\phi_{+}(x)\, ,
\label{SG+}\\
{\cal H}^{-} & = &  {u_{-} \over 2}[(\partial_x \theta _{-})^{2} 
+ (\partial_x \phi_{-}(x))^2] \nonumber\\
&+& {\cal J}_{\bot}\cos \sqrt{\frac{2\pi}{K_{-}}}\theta_{-}(x)\, ,
\label{SG-}\\
{\cal H}^{\pm}  & = & - h 
\cos\sqrt{\frac{ \pi}{2K_{-}}}\theta_{-}(x) 
\cos\sqrt{\frac{ \pi}{2K_{+}}}\theta_{+}(x)\, 
\label{SGint} 
\end{eqnarray}


Here
\be\label{Kpm}
K_{\pm} \simeq  K \left ( 1 \mp \frac{J_{\perp}}{2\pi J}
\frac{2K-1}{\sin(\pi/2K)}\right)\, ,
\ee
${\cal J}_{\bot}\sim  J_{\bot}$, ${ h} \sim { h^{ext}}$ and $u_{\pm}$ are the velocities of the symmetric and antisymmetric modes.

In deriving (\ref{EffectiveHamiltonian}), several terms which are
strongly irrelevant in our case of a ladder with ferromagnetic legs
and applied transverse magnetic field, were omitted. For details of
the full Hamiltonian we refer the reader to \cite{Vekua}.

We note that at ${h^{ext}}=0$ the effective theory of the original ladder
model is given by two decoupled quantum sine-Gordon models which describe,
respectively, the symmetric and antisymmetric degrees of freedom.
The infrared properties of the antisymmetric field are governed by the 
{\em strongly relevant} operator 
${\cal J}_{\bot}\cos \sqrt{2\pi/K_{-}}\theta_{-}$ with the scaling dimension 
$ (2 K_{-})^{-1} \leq 1/2$. Therefore, the {\em antisymmetric 
sector is gapped at arbitrary} $J_{\bot} \neq 0$. Fluctuations of the field
$\theta_{-}(x)$ are completely suppressed and the field $\theta_{-}$ gets 
ordered with expectation values
\begin{equation}  
\label{ORDERFIELDS}
\langle\theta_{-}\rangle =\left\{ 
\begin{array}{ll}
\sqrt{K_- \pi/2} \hskip0.5cm \textrm{ at ${\cal J}_{\bot}>0$}  \\ 
0                \hskip1.8cm \textrm{ at ${\cal J}_{\bot} < 0$}
\end{array}
\right. \, .
\end{equation}

The infrared properties of the symmetric field are governed by the
{\em marginal} operator $M_{+}\cos \sqrt{8\pi K_{+}}\phi_{+}$. As
shown in \cite{Vekua}, the symmetric mode remains gapless for
ferromagnetic interleg exchange and arbitrary $0\leq \Delta \leq1$, while in
the case of antiferromagnetic interleg exchange it is gapless in a
finite regime in $\Delta,J_{\bot}$ parameter space given approximately
by $\gamma_{1}J_{\bot}/J\leq \Delta\leq 1-\gamma_{2}J_{\bot}/J$. Here
$\gamma_{1}$ and $\gamma_{2}$ are positive constants (more rigorously
smooth functions of the anisotropy) of the order of unity. Following
Schulz \cite{Schulz1} who has discussed a similar phase in the context
of the spin $S=1$ chain we denoted this phase as {\em spin liquid XY1}
phase for the spin ladder. At $J_{\bot}>0$, outside of this regime,
the symmetric mode is also gapped and the field $\phi_{+}$ gets
ordered and pinned in one of its possible minima. We denoted the fully
gapped phase, realized in the case of antiferromagnetic interleg
exchange as rung-singlet phase. In this phase spins on the same rung
tend to form a singlet and the ground state corresponds to the state
with a singlet pair on each rung. The ground state phase diagram
of the system at ${h^{ext}}=0$ is given schematically in
Fig.~\ref{fig:ladder}.

Below we study the ground state phase diagram of the two-leg
ladder with {\em anisotropic ferromagnetic} legs in the presence of
a magnetic field.

\section{The spin ladder in the presence of an in-plane magnetic field}

In this Section we consider the effect of a transverse magnetic field
on the ground state phase diagram of the spin-liquid $XY1$ phase.
Since the magnetic field breaks the in-plane rotational symmetry of
the $XY1$ phase, it is clear that it will eliminate the gapless $XY1$
phase and that the system will acquire a gap in the excitation
spectrum for arbitrary small ${h^{ext}}$. Both fields (symmetric and
antisymmetric) will be pinned in their respective minima. In this
case the very first direct approach to study the ground state phase
diagram of the model is to use the quasiclassical Ginzburg-Landau type
analysis. This approach does not cover the limiting case $\Delta=1$ where 
bosonization does not apply. The $\Delta=1$ case will be discussed in section 
IV.

\subsection{Quasiclassical Ginzburg-Landau analysis}

In the presence of an in-plane magnetic field the effective Hamiltonian 
(\ref{EffectiveHamiltonian}) reduces to two free Gaussian fields 
coupled by the following effective potential
\be\label{EffPot}
{\cal U}_{eff}\left( \Theta_{-},\Theta_{+}\right) = 
{\cal J}_{\bot}\cos 2\Theta_{-} - h \cos\Theta_{-}\cos\Theta_{+}\, , 
\ee
where $\Theta _\pm=\sqrt{\pi/ 2K_{\pm}} \theta_{\pm}$. To find the
vacuum expectation values of the pinned field we search for the minima
of the effective potential ${\cal U}_{eff}$ with respect to
$\Theta_{-}$ and $\Theta_{+}$. The straightforward analysis gives the
following sets of vacua:
\begin{itemize}
\item{$ h>  h_{c}=4 {\cal J}_{\bot}$}
\end{itemize}
\begin{eqnarray}
\label{vacua1a}
{\rm I.}\hskip0.3cm \Theta_{+} &=& \pi \, ,\,\Theta_{-} = \pi\, , 
~~ ({\rm mod~} 2\pi )\nonumber\\
{\rm II.}\hskip0.3cm \Theta_{+} &=& 0  \, ,\, \Theta_{-} = 0\, ,  
~~ ({\rm mod~} 2\pi )
\end{eqnarray}
\begin{itemize}
\item{$h< h_{c} = 4 {\cal J}_{\bot}$}
\end{itemize}
\begin{eqnarray}
{\rm I.}\hskip0.3cm \Theta_+&=&\pi \,\,\Theta_{-} =\pi \pm \vartheta_{0}\, , 
~~ ({\rm mod~} 2\pi )\nonumber\\
{\rm II.}\hskip0.3cm \Theta_+&=&0,\,\, \Theta_-= \pm \vartheta_{0}    \, , 
~~({\rm mod~} 2\pi) 
\label{vacua2a}
\end{eqnarray}
where 
$$ \vartheta_{0}=\arccos( h/4 {\cal J}_{\bot}). $$ 

At ${h}=0$ only the antisymmetric field is gapped and its set
of available vacua is given by $\Theta_{-}=\pi/2, ({\rm mod~} \pi )$.
At ${h} \neq 0$ the symmetric field gets pinned and the set of
possible vacua of the symmetric field does not change with field. On
the other hand, at arbitrary $0 < h < 4 {\cal J}_{\bot}$, each minimum
in the antisymmetric sector splits into two degenerate minima (see
Fig.~\ref{fig:vacua}). At $h \rightarrow {h}_{c}$, $\vartheta_{0}
\rightarrow \pi/2$ the split minima join each other and form a new set
of possible vacua for the antisymmetric field which is fixed for
arbitrary $ h> h_{c}$. At the critical point the effective potential
transforms into the $(\Theta_{-})^{4}$ potential, which is common in
describing the Ising universality class. Since the location of the
minima in the symmetric sector does not change, we conclude that the
transition involves only the antisymmetric sector.


\begin{figure}[tb]
\includegraphics[width=80mm]{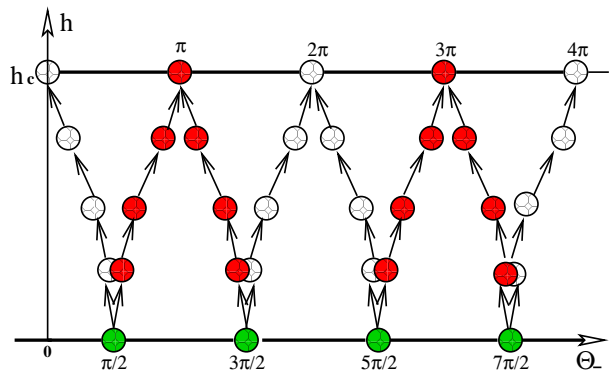}
\caption{\label{fig:vacua} Evolution of the set of minima of the 
antisymmetric field with increase of the applied transverse magnetic field.}
\end{figure}


An analogous quasiclassical analysis was carried out by Fabrizio et
al.  \cite{Fabrizio} in connection with the ionic Hubbard model. There
two transitions were identified for finite values of alternating
on-site energies. We note as an important difference that assuming a
vanishing order parameter along the direction of an infinitesimally small
applied magnetic field would be self-contradictory in our case,
because here no appropriate marginal operators are present.

Note that in the case of ferromagnetic interleg exchange
($J_{\bot}<0$), the set of minima of the effective potential for
arbitrary ${h^{ext}} \neq 0$ is given by the set (\ref{vacua1a}).
Therefore, no transitions with increasing field are expected in this
case.

\subsection{Phase diagram}

The quasi-classical analysis performed above allows to sketch
qualitatively the ground state phase diagram of the model under
consideration.

As soon as we switch on an infinitesimally small magnetic field in $X$
direction, breaking the in-plane rotation symmetry, the $XY1$ phase is
expelled, the system becomes gapped and acquires a finite order
parameter. Using the bosonized expressions for the smooth parts of the
spin operators Eqs.~(\ref{bosforSx})-(\ref{bosforSy}), expressed in
terms of the symmetric and antisymmetric fields and the vacuum
expectation values of the corresponding fields
(\ref{vacua1a})-(\ref{vacua2a}) one easily obtains that at $0<h<h_{c}$
the spin ladder acquires
\begin{itemize}
\item{uniform magnetization in the direction of applied field $M_{x}$}
\end{itemize}
\begin{eqnarray}
M_{x} &=&\left< S_{j,1}^{x}(x)\right>  = \left< S_{j,2}^{x}(x)\right>
\nonumber\\
&\simeq &\left< \cos\Theta_{+}\right>\left<\cos \Theta_{-}\right>=   
h/4{\cal J}_{\bot} 
\end{eqnarray}
\begin{itemize}
\item{opposite magnetization of legs in the in-plane direction
perpendicular to the field:}
\end{itemize}
\begin{eqnarray}
M_{y}&=&\left< S_{j,1}^{y}(x)\right>  = -\left< S_{j,2}^{y}(x)\right> 
\nonumber\\
&\simeq&
\left< \cos\Theta_{+}\right>\left<\sin \Theta_{-}\right> = 
\sqrt{1-\left(h/4{\cal J}_{\bot}\right)^{2}}\, . 
\end{eqnarray}
This phase we denote as the "stripe-ferromagnetic" phase.

When the magnetic field exceeds the critical value $h>h_{c}$, the new
set of vacua (\ref{vacua1a}) is reached and the system passes into the
ferromagnetic phase, where

\begin{equation}\label{noneel}
M_{x} \simeq \left < \cos \Theta_- \right> = 1 \,{\rm and}\,\quad
M_{y} \simeq \left < \sin \Theta_- \right> = 0\, .
\end{equation}

Thus, the quasiclassical analysis of the ground state phase diagram of
the $XY1$ phase in the presence of a transverse magnetic field shows a
{\em phase transition} from the stripe-ferromagnetic to the
ferromagnetic phase (we denote by 'ferromagnetic phase' the phase with
the magnetization parallel to the external field as only nonvanishing
order parameter).  This analysis also indicates, that the transition
belongs to the Ising universality class.
\begin{figure}[tb]
\includegraphics[width=60mm]{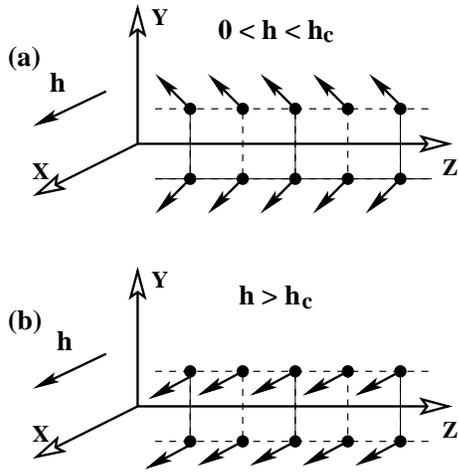}
\caption{\label{fig:crossover}Crossover from stripe-ferromagnetic (a) to ferromagnetic (b) states.}
\end{figure}

From the above it is also clear that in the case of ferromagnetic interleg 
exchange (${\cal J}_{\bot}<0$) the ferromagnetic phase is realized 
for arbitrary $h \neq 0$: This is so, since in this case the vacua of two 
terms in (\ref{EffPot}) do not exclude each other.

\subsection{Double sine-Gordon analysis} 

According to the quasiclassical analysis the magnetic field induced
transition in the $XY1$ phase from the partly polarized
stripe-ferromagnetic phase into the ferromagnetic phase takes place
along the critical line $h_{c} = 4{\cal J}_{\bot}$ and belongs to the
Ising universality class.  In this subsection we investigate the
stability against quantum fluctuations of the results obtained so far.
For this purpose we decouple the initial theory of interacting quantum
multi-frequency sine-Gordon fields (\ref{EffectiveHamiltonian}) to get

\begin{eqnarray}\label{EffectiveHamiltonianMF}
{\cal H}_{MF}& =& {\cal H}^{+} + {\cal H}^{-}\, ,\\   
{\cal H}^{+} & =& {u_{+} \over 2} [(\partial_x \theta_{+})^{2} + 
(\partial_x \phi_{+})^2] \nonumber\\
& -& h_{+} \cos\beta_{+}\theta_{+}(x)\, ,
\label{SSG+}\\
{\cal H}^{-}  &=&   {u_{-} \over 2}[(\partial_x \theta _{-})^{2} 
+ (\partial_x \phi_{-}(x))^2] \nonumber\\
&-& h_{-} \cos\beta_{-}\theta_{-}(x) + 
{\cal J}_{\bot}\cos2\beta_{-}\theta_{-}(x)\, ,
\label{SSG-}
\end{eqnarray}
where
\be\label{MFbeta}
\beta_{\pm} = \sqrt{\pi/2K_{\pm}} 
\ee
and
\be\label{MFmasses}
h_{\pm} =  h\langle \cos\sqrt{\pi/2K_{+}}\theta_{\mp} \rangle
\ee

Thus the symmetric sector is described by the ordinary sine-Gordon theory with 
$\beta_{+}^2 \leq \pi/2 $ for $0\leq \Delta \leq1$, and therefore with 
a strongly relevant massive term. 
On the other hand, in the antisymmetric sector we arrive at the double 
frequency sine-Gordon 
model with $\beta_{-}^2 \leq \pi/2$  and therefore strongly relevant 
basic and double-field operators.  

In the double frequency sine-Gordon model the quantum phase transition
in the ground state takes place when the vacuum configurations of the
two {\em cosines} compete with each other \cite{Delfino}, corresponding to the
crossover from double well to single well potential in quasiclassical 
analysis. 
One easily verifies that this criterion can be applied
analogously to the case of antiferromagnetic rung exchange
($J_{\bot}>0$). The dimensional arguments based on equating physical
masses produced by the two cosine terms separately is usually used to
define the critical line:
\begin{equation}
\left\{ \begin{array}{lll}
m_h& = & (h)^{1/\left(2-dim[h]\right)}\\
m_{J_{\bot}}&=&   J_{\bot}^{1/\left(2-dim[J_{\bot}]\right)}
\end{array}
\right.
\end{equation}
Equating these two masses
we obtain the following expression for the critical line
external magnetic field vs. interchain coupling:
\begin{equation}
\label{criticalline}
h_{cr}=\eta J_{\bot}^{\mu(\Delta)}\, .
\end{equation}
where
\begin{equation}
\label{mu}
\mu={2-dim[h]\over 2-dim[J_{\bot}]}=(8K^2-K)/2K_+(4K_--1)
\end{equation}
and $\eta$ is some numerical constant of the order of unity.  The self
consistency of the mean field separation insures that (\ref{mu})
follows from equating masses produced by magnetic field and interchain
couplings separately, before the mean field separation.  From
(\ref{mu}) we see $h_{cr}\simeq J_{\bot}$ in the limit $K\to \infty$,
i.e. for the single chain ferromagnetic instability point. This will
be seen to be consistent with the large rung coupling as well as with spin wave
analysis (see respectively eqs. (\ref{Ferroline}) and (\ref{ferroboundary})).

Finally we want to mention that the operator product expansion of the
two lowest frequencies of the sine-Gordon theory for the range of
anisotropy parameter $0 \leq \Delta \leq 1$ does not close,
i.e. higher relevant harmonics are generated in the RG procedure. In
some situations it is known that higher harmonics can destabilize
second order phase transitions and make it weakly first
order\cite{Palla}. This happens when generated harmonics introduce new
minima (e.g. the dashed minimum in Fig. (\ref{fig:doublewell})) which
at the phase transition point coexists with the minima before the
transition.

\begin{figure}[tb]
\includegraphics[width=60mm]{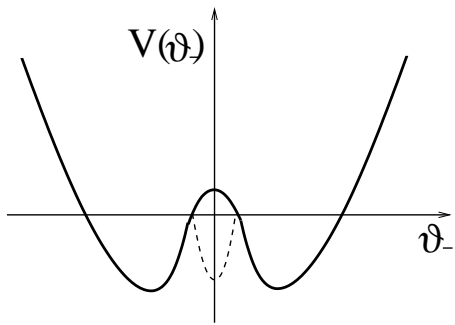}
\caption{\label{fig:doublewell} Appearance of an additional minimum (dashed 
line) which could change the second order phase transition to first order, 
depending on the sign of the higher order harmonic generated.}
\end{figure}

For example in the case of only the third harmonic retained, the
position of the new minimum depends on the sign of the generated
harmonics. We have followed the sign of the generated harmonic along
the one loop RG flows and checked that in our case there is no
indication of an upsetting of the second order phase transition.  The
new minima get pinned in between the minima of the first two
harmonics; they thus may shift the critical line (\ref{criticalline})
to larger values of the magnetic field but otherwise do not introduce
any qualitatively new effects.


\section{Large rung coupling results}

In this section we consider the effect of an applied magnetic field on
the ground state phase diagram of the two-leg ladder system with
ferromagnetic legs (\ref{Hamiltonian}) in the limiting case of strong
rung coupling $J_{\bot}\gg J$. In this limit it is convenient to
discuss the model by representing the {\em site}-spin algebra in terms
of {\em on-bond}-spin operators \cite{SachdevBhatt}. In particular, in
the case of isotropic interleg exchange the ladder can be mapped onto
a single $S=1/2$ chain \cite{Chaboussant2,Totsuka,Mila}. For
completeness we briefly discuss the mapping here: A given
rung may be in the singlet or in the triplet state with
energies given by
$$ E_{t,\pm}={J_{\bot}\over 4}\ \pm {h^{ext}},\,\, E_{t,0}={J_{\bot}\over
4},\,\,E_s=-{3J_{\bot}\over 4}.  $$
At ${h^{ext}} \leq J_{\bot}$, one component of the triplet becomes
closer to the singlet ground state such that for a sufficiently strong
magnetic field we have a situation where the singlet and the $S_{z} = +1$
component of the triplet form a new effective spin 1/2 system. One can
easily project the original ladder Hamiltonian (\ref{Hamiltonian}) on
the new singlet-triplet subspace
\begin{eqnarray} |\Uparrow\rangle & \equiv& |t^+\rangle =
|\uparrow \uparrow \rangle\\
|\Downarrow\rangle &\equiv&|s\rangle =  \frac1{\sqrt2}[|\uparrow\downarrow\rangle -|\downarrow\uparrow\rangle]
 \nonumber \end{eqnarray}
This leads to the definition of the effective spin 1/2 operators
\begin{eqnarray} S^+_{n,\alpha=1,2} &=& (-1)^{\alpha}
\frac1{\sqrt2}\tau^{+}_{n} \label{tau+} \\ S^z_{n,\alpha=1,2} &=&
\frac{1}{4} [I + 2\tau^{z}_{n}] \label{tauz} \end{eqnarray}
When expressed in terms of the effective spin operators
(\ref{tau+})-(\ref{tauz}), the original Hamiltonian
(\ref{Hamiltonian}) becomes
\begin{eqnarray}\label{EffectiveHamiltonian2} H_{\text{eff}} & = & -J
\sum_i [\frac{1}{2}\tau^{z}_{i}\tau^{z}_{i+1} + \tau^{y}_{i}
\tau^{y}_{i+1} +\Delta \tau^{x}_{i} \tau^{x}_{i+1}] \nonumber \\
 & & - {h}_{eff} \sum_i \tau^{z}_{i} + {\mbox Constant}, 
\label{eq:hameffective}
\end{eqnarray}
where the effective magnetic field
\begin{equation}\label{Heffective}
{h}_{eff} = {h^{ext}} - J_\perp - \frac{1}{2}J.
\end{equation}
Note that in writing (\ref{EffectiveHamiltonian2}) for convenience we have 
exchanged $X$ and $Z$ axis in effective spin space.
\begin{figure}[tb]
\includegraphics[width=80mm]{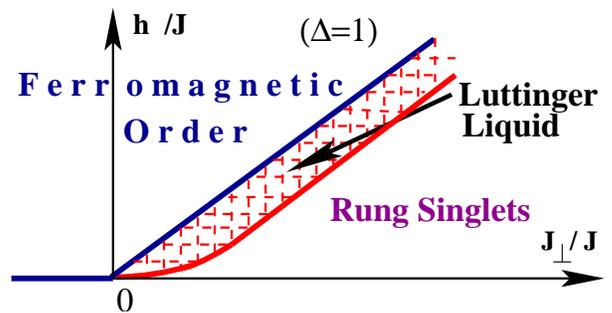}
\caption{\label{fig:Fig4}Phase diagram of a spin ladder with the SU(2) 
symmetric ferromagnetic legs in a uniform magnetic field.}
\end{figure}

The Hamiltonian (\ref{EffectiveHamiltonian2}) describes a spin 1/2
{\em fully anisotropic} $XYZ$ chain in an effective magnetic field.
There exists a special case, $\Delta=1$, which allows for rigorous
analysis. In this case the effective problem reduces to the theory of
the $XXZ$ chain with a \emph{fixed ferromagnetic} $XY$ anisotropy of
$1/2$ in an effective magnetic field ${h}_{eff}$. The gapped
phase at $ {h}_{eff} <{ h}^{c1}_{eff}$ for the ladder
corresponds to the fully saturated magnetization phase for the
effective spin chain pointing in the direction opposite to the applied field, 
whereas the massless phase for the
ladder corresponds to the finite magnetization phase of the effective
spin-1/2 chain \cite{Chaboussant2}. The critical field ${h}
^{c2}_{eff}$ where the ladder is fully magnetized corresponds to the
fully magnetized phase of the effective spin chain pointing along the direction of applied field. 
From the exact ground state phase diagram of the anisotropic $XXZ$ chain in
a magnetic field \cite{Takahashi} using (\ref{Heffective}) we get
that the {\em isotropic ferromagnetic ladder} in a magnetic field
shows two second order phase transitions: at
\be {h^{ext}}_{c1}= J_{\bot}-J \ee
a transition occurs from the rung dimer to a Luttinger liquid phase
and at
\be \label{Ferroline} {h^{ext}}_{c2}=J_{\bot} \ee
a transition from a Luttinger liquid phase into the fully polarized phase.

The transition from the rung-singlet phase into the Luttinger liquid
phase in the case of the {\em isotropic antiferromagnetic ladder} was
studied in detail in several recent publications
\cite{Totsuka,Usami,Mila,GiamarchiTsvelik}. It was shown that in the
case of the gapped rung-singlet phase the magnetization appears only
at a finite critical value of the magnetic field equal to the spin
gap. Since this behavior is generic for isotropic systems with spin
gap \cite{JNW}, and the gap in the ladder system is governed by
$J_{\bot}$ we conclude that the rung-singlet to Luttinger liquid phase
transition line smoothly reaches zero at $J_{\bot} \rightarrow 0$ (see
Fig.\ref{fig:Fig4}).

Note that this large rung coupling analysis reveals that the phase
transitions in the antiferromagnetically coupled ladder with
ferromagnetic legs in uniform magnetic field are connected with those
in a ladder with antiferromagnetic legs but in staggered magnetic
field \cite{Wang2}. In both cases the magnetic field tries to promote
triplets on rungs, while the antiferromagnetically coupled ladder
supports on-rung singlets.

Away from the isotropic point $\Delta=1$ the effective Hamiltonian
(\ref{EffectiveHamiltonian2}) describes the fully anisotropic
ferromagnetic $XYZ$ chain in a magnetic field that is directed
perpendicular to the easy axes. For the
particular value of magnetic field $\mathrm{h}_{eff}=0$, the effective
$XYZ$ chain is long range ordered in $Y-$direction, corresponding to
the original ladder system being ordered in the direction
perpendicular to the applied magnetic field with opposite
magnetization on the legs (stripe-ferromagnetic phase). For larger values of
the effective field it is clear that this striped ferromagnetic order
will be replaced either by the rung singlet phase or the phase with
only one order parameter - magnetization along the applied field.
Thus we obtain the result that the phases appearing in large rung
coupling and in weak rung coupling are identical. On general grounds
one expects an Ising transition to take place at critical strengths of
the magnetic field also for large couplings. Thus for $\Delta \neq 1$
the Luttinger liquid phase will be replaced by the striped
ferromagnetic phase (as in the weak coupling limit), and the
transitions become of Ising type due to the reduced symmetry
(Fig.~\ref{fig:phase}). 

In order to determine the nature of the
transition from rung singlets to the striped ferromagnetic state in
the weak coupling limit we can use the fact that in the rung singlet
phase the operator
$$ J_{\bot}\cos{\sqrt {8 \pi K_{+}}\phi_{+}(x)}$$
is relevant and the ground state consists of nonmagnetic singlets
situated along the rungs of the ladder. On the other hand, while the
in-plane magnetic field couples to the dual fields (disorder
operators) we expect an Ising phase transition to take place with the
appearance of the magnetization perpendicular to the applied field
(stripe-FM) as dominant order parameter.
\begin{figure}[tb] 
\includegraphics[width=80mm]{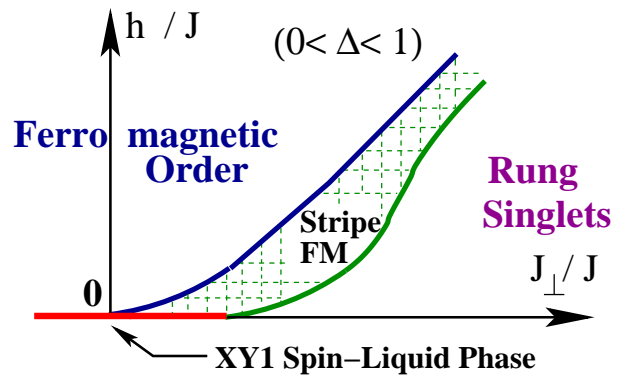}
\caption{\label{fig:phase}Phase diagram of a ladder with anisotropic
ferromagnetic legs in transverse magnetic field.}  
\end{figure}

\section {Conclusions}

We have investigated the phase diagram of the $S=1/2$ ladder with
ferromagnetic legs under the influence of a uniform magnetic field
breaking the in-plane rotational symmetry. In case of
antiferromagnetic coupling between legs we identified two phase
transitions in the plane of magnetic field vs interchain coupling. We
have extended our analysis to the strong rung coupling limit and have
identified a Luttinger liquid phase which replaces the stripe-
ferromagnetic phase in the case of $SU(2)$ symmetric legs. The
phase transition line in the case of $SU(2)$ ferromagnetic legs was
confirmed also by the spin wave calculation starting from the
saturated limit.


\medskip 

\section{Acknowledgements} TV is grateful to A. A. Nersesyan for
interesting discussions. GIJ acknowledges support by the SCOPES grant N
7GEPJ62379.  This work was supported in part by the DFG-Graduiertenkolleg
No. 282, ''Quantum Field Theory Methods in Particle Physics,
Gravitational Physics, and Statistical Physics''.

\medskip



\appendix

\section{Ferromagnetic instability}

In this appendix we use the spin wave approach to determine the
critical line corresponding to the {\em ferromagnetic instability} for
the case of $SU(2)$ symmetric legs. We refer the reader to the
appendix of (\onlinecite{Vekua}) where we have considered in detail
the analogous analysis of a ladder without magnetic field. In the case
of $SU(2)$ symmetric legs it is straightforward to add the magnetic
field term, since it couples to the diagonal operator. In this case
the two sets of spin wave excitation frequencies \cite{Vekua}
read:
\begin{eqnarray} \omega^{-}(q) & = & {h^{ext}}-{J_{\bot}\over 2}-J
\cos q-{J_{\bot}\over 2} \\ \omega^{+}(q) & = &
{h^{ext}}-{J_{\bot}\over 2}-J
                                             \cos q+{J_{\bot}\over 2} \, .
\end{eqnarray}
For $J_{\bot}>0$ we have
$$ \omega^{-}(q) < \omega^{+}(q) $$
and from the instability condition $\omega^{-}(q=0)=0$ we obtain
\be 
\label{ferroboundary}
{h^{ext}}=J_{\bot}
\ee



\begin{thebibliography}{99}

\bibitem{RiceDagotto}{For a review see E. Dagotto, Rep. Prog. Phys. {\bf 62}, 
1525 (1999); E. Dagotto and T.M. Rice, Science {\bf 271}, 618 (1996).}

\bibitem{Chaboussant1} G. Chaboussant and P. A. Crowell, L. P. L\'evy, 
O. Piovesana, A. Madouri,  and D. Mailly, Phys. Rev. B {\bf  55}, 3046 (1997).

\bibitem{Chaboussant2} G. Chaboussant, M.-H. Julien, Y. Fagot-Revurat,
M. Hanson, L.P. L\'evy, C. Berthier, M. Horvatic, and O. Piovesana,  
Europ. Phys. J. B {\bf 6}, 167 (1998).

\bibitem{Chaboussant3} G. Chaboussant,  Y. Fagot-Revurat,  M.-H. Julien, 
M. E. Hanson,  C. Berthier,  M. Horvatic, L. P. L\'evy, and O. Piovesana, 
Phys. Rev. Lett. {\bf 80}, 2713 (1998).

\bibitem{Exp4} D. Arcon, A. Lappas, S. Margadonna, K. Prassides, E. Ribera, 
J. Veciana, C. Rovira, R. T. Henriques, and M. Almeida, Phys. Rev. B {\bf 60}, 
4191 (1999). 

\bibitem{Exp5} H. Mayaffre, M. Horvatic, C. Berthier, M.-H. Julien, 
P. S\'egransan, L. L\'evy, and O. Piovesana, Phys. Rev. Lett. {\bf 85}, 4795 
(2000).

\bibitem{Exp6} B. C. Watson, V. N. Kotov, M. W. Meisel, D. W. Hall, 
G. E. Granroth, W. T. Montfrooij, S. E. Nagler, D. A. Jensen, R. Backov, M.
A. Petruska, G. E. Fanucci, and D. R. Talham, Phys. Rev. Lett. {\bf 86}, 5168 
(2001).

\bibitem{ChitraGiamarchi} R. Chitra and T. Giamarchi, Phys. Rev. B 
{\bf  55}, 5816 (1997).

\bibitem{Honecker} D. C. Cabra, A. Honecker, and P. Pujol, Phys. Rev. Lett. 
{\bf  79}, 5126 (1997);\, {\it ibid}  Phys. Rev. B {\bf  58}, 6241 (1998).

\bibitem{Totsuka} K. Totsuka,  Phys. Rev. B {\bf  57} 3454 (1998).

\bibitem{Usami} M. Usami and S. I. Suga, Phys. Rev. B {\bf  58}, 14401 (1998.)

\bibitem{GiamarchiTsvelik} T. Giamarchi and A. M. Tsvelik, Phys. Rev. B 
{\bf  59}, 11398 (1999).

\bibitem{Mikeska1} M. Hagiwara, H. A. Katori, U. Schollw{\"o}ck, and 
H.-J. Mikeska, Phys. Rev. B {\bf  62}, 1051 (2000).

\bibitem{Wang1} X. Wang and L. Yu, Phys. Rev. Lett. {\bf  84}, 5399 (2000).

\bibitem{Haas}S. Wessel, M. Olshanii, and S. Haas, Phys. Rev. Lett. 
{\bf  87}, 206407 (2001)

\bibitem{Wang2} Y.-J. Wang, F. H. L. Essler, M. Fabrizio, and A. A. Nersesyan, 
Phys. Rev. {\bf  B 66}, 024412 (2002).

\bibitem{Schulz1}{H.J. Schulz, Phys. Rev. B {\bf 34}, 6372, (1986).}

\bibitem{KolezhukMikeska} A. K. Kolezhuk and H.-J. Mikeska, Int. J. Mod. Phys. 
{\bf 12}, 2325 (1998).

\bibitem{Vekua} T. Vekua, G.I. Japaridze, and H.-J. Mikeska, Phys. Rev. B 
{\bf 67}, 064419, (2003). 

\bibitem{Nersesyan96} D.G. Shelton, A.A. Nersesyan and A.M. Tsvelik, 
Phys. Rev. B {\bf 53}, 8521 (1996).

\bibitem{JNW} G. I. Japaridze, A.A. Nersesyan and P.B. Wiegmann, Nucl. Phys. 
B 230 (FS 4), 511-547 (1984). 

\bibitem{Japaridze} G. I. Japaridze and A. A. Nersesyan, Pis'ma Zh. 
Eksp. Teor. Fiz. {\bf 27}, 356 (1978) [Sov. Phys. JETP Lett. {\bf 27}, 
334  (1978)].

\bibitem{Pokrovsky} V.L. Pokrovsky and A.L. Talapov, Phys. Rev. Lett. 
{\bf 42}, 65 (1979).

\bibitem{Mikeska} H.-J. Mikeska, J. Phys. C: Solid St. Phys. {\bf 11}, 
L29, (1978); H.-J. Mikeska and M. Steiner, Advances in Physics, 
{\bf 40}, 191, (1991).

\bibitem{Hieida} Y. Hieida, K. Okunishi, and Y. Akutsu, Phys. Rev. B 
{\bf  64}, 224422 (2001).

\bibitem{Ovchinnikov} D.V. Dmitriev, V.Ya. Krivnov,and A.A. Ovchinnikov, Phys. 
Rev.B {\bf 65}, 172409 (2002); D.V. Dmitriev, V.Ya. Krivnov, A.A. Ovchinnikov, 
and A. Langari, JETP {\bf 95}, 538 (2002).

\bibitem{Essler03}  J-S. Caux, F.H.L. Essler, and U. L{\"o}w, 
Phys. Rev. B {\bf 68}, 134431 (2003).

\bibitem{DuttaSen}A. Dutta and D. Sen, Phys. Rev. B {\bf  67}, 094435 (2002).
\bibitem{Fabrizio} M. Fabrizio, A.O. Gogolin, and A.A. Nersesyan, Phys. Rev. 
Lett. {\bf 83}, 2014, (1999); {\em ibid} Nucl.Phys. B {\bf 580}, 647 (2000).

\bibitem{Delfino} G. Delfino and G. Mussardo, Nucl. Phys. B {\bf  516}, 675 
(1998).

\bibitem{Palla} Z. Bajnok, L. Palla, G. Takacs, F. Wagner, Nucl. Phys. B 
{\bf  601} 503, (2001).

\bibitem{SachdevBhatt} S. Sachdev and R.N. Bhatt,  Phys. Rev. B {\bf 41}, 9323 
(1990).

\bibitem{Mila} F. Mila, Eur. Phys. J. B {\bf 6}, 201 (1998). 

\bibitem{Takahashi} M. Takahashi, { \it Thermodynamics of one-dimensional 
solvable models}, Cambridge University Press (1999), Chapter IV.


\end{thebibliography}
\end{document}